\pdfoutput=1
\documentclass[%
aps,
prl,
amsmath,amssymb,
reprint,
showpacs,
superscriptaddress,
]{revtex4-1}
\usepackage[]{graphicx} 
\usepackage[]{pdfpages}
\def\U#1{{%
\def\O{\mbox{O}}
\def\u{\mbox{u}}
\mathcode`\u=\mu
\mathcode`\O=\Omega
\mathrm{#1}}}
\def\vct#1{{\mathchoice{\mbox{\boldmath$#1$}}{\mbox{\boldmath$#1$}}%
  {\mbox{\scriptsize\boldmath$#1$}}{\mbox{\scriptsize\boldmath$#1$}}}}
\def\sub#1{_{\mbox{\scriptsize#1}}}
\def\Re{\mathop{\mathrm{Re}}}
\def\Im{\mathop{\mathrm{Im}}}
\begin{document}
\preprint{Ver. 1.6}
\title{Frequency-Independent Response of Self-Complementary Checkerboard Screens}
\author{Yoshiro Urade}
\email{urade@giga.kuee.kyoto-u.ac.jp}
\affiliation{Department of Electronic Science and Engineering, Kyoto University, Kyoto 615-8510, Japan}
\author{Yosuke Nakata}
\affiliation{Center for Energy and Environmental Science, Shinshu University, 4-17-1 Wakasato, Nagano 380-8553, Japan}
\author{Toshihiro Nakanishi}
\affiliation{Department of Electronic Science and Engineering, Kyoto University, Kyoto 615-8510, Japan}
\author{Masao Kitano}
\email{kitano@kuee.kyoto-u.ac.jp}
\affiliation{Department of Electronic Science and Engineering, Kyoto University, Kyoto 615-8510, Japan}
\date{\today}
\begin{abstract}
This research resolves a long-standing problem on the electromagnetic response of self-complementary metallic screens with checkerboardlike geometry.
Although Babinet's principle implies that they show a frequency-independent response, this unusual characteristic has not been observed yet due to the singularities of the metallic point contacts in the checkerboard geometry.
 We overcome this difficulty by replacing the point contacts with resistive sheets.
 The proposed structure is prepared and characterized by terahertz time-domain spectroscopy.
 It is experimentally confirmed that the resistive checkerboard structures exhibit a flat transmission spectrum over 0.1--1.1~THz.
 It is also demonstrated that self-complementarity can eliminate even the frequency-dependent transmission characteristics of resonant metamaterials. 
\end{abstract}
\maketitle
%
%
Duality is one of the key concepts in physics and engineering, as exemplified by electromagnetic duality~\cite{Bliokh2013}, $T$ duality in string theory~\cite{Alvarez1995}, and the duality of electrical circuits~\cite{Desoer1969basic}. 
It relates two seemingly different systems or quantities, and sometimes helps us to indirectly gain physical insight into intractable problems.
A system is said to be self-dual if it coincides with its own dual.
Self-duality is symmetry with respect to duality transformations.
Problems with self-duality often have simple analytical solutions due to the constraints imposed by their symmetry.
These results are universal and do not depend on the details of the problems.
For example, self-dual symmetry has been utilized to determine critical temperatures of two-dimensional Ising models~(Kramers-Wannier duality)~\cite{Kramers1941}.
Moreover, there are examples in electromagnetic systems ranging from dc to radio frequency: evaluating the effective conductivity of two-phase composite media~(Keller-Dykhne duality)~\cite{Keller1964,*Dykhne1970}, obtaining constant-resistance electrical circuits~\cite{LinSelfDual}, and making broadband antennas~\cite{Mushiakebook}. Recently, self-dual symmetry has been applied to the design of metamaterials with zero backscattering~\cite{Lindell2009}.
\begin{figure}[b]
 \includegraphics[width=7.5cm]{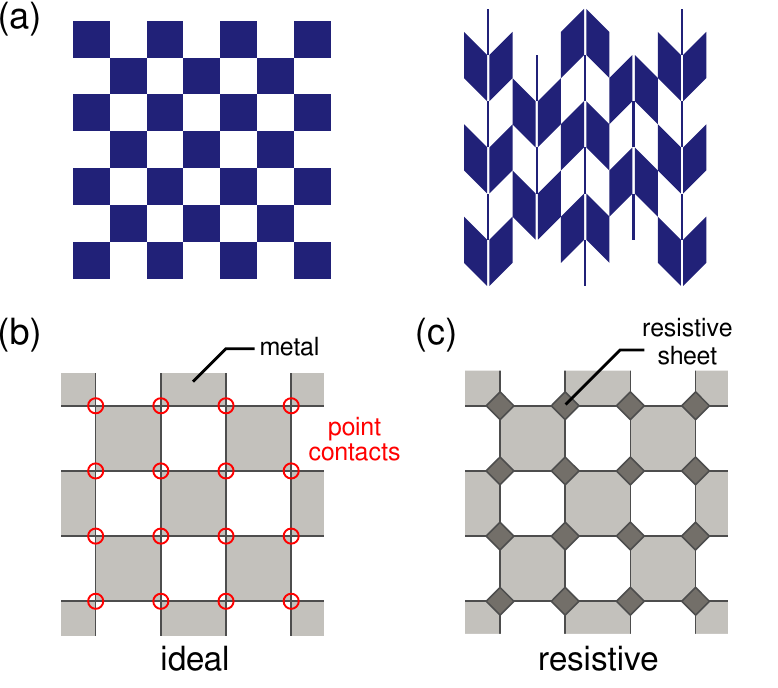}
 \caption{\label{fig:res_check}(color online). (a)~Examples of planar patterns  with self-complementarity. (b)~The ideal metallic checkerboard structure with self-complementarity, and (c)~the resistive checkerboard structure.}
\end{figure}

Let us now focus our attention on the duality and self-duality of planar structures composed of two elements, which can be represented by two-tone patterns. Interchange of these two colors is considered to be a duality transformation. If patterns are invariant under the interchange of colors, as shown in Fig.~\ref{fig:res_check}(a), they are self-dual. This type of symmetry is referred to as color symmetry~\cite{Senechal1988} or self-complementarity, and it is often found in designs of traditional garments and in the impressive works of the graphic artist M.\,C.~Escher.

In optics, there is a well-known duality relationship called Babinet's principle~\cite{JacksonBook}, which relates the fields scattered by a metallic screen with those scattered by its complementary screen, which is obtained by interchanging the areas of metal and the holes.
With Babinet's principle, it has been shown theoretically and experimentally that an antenna with a self-complementary shape exhibits a frequency-independent input impedance~\cite{Mushiakebook,BalanisAntenna}.
Similarly, it also predicts that self-complementary metallic screens exhibit frequency-independent responses as described below.
Here, we will consider the scattering problems of self-complementary metallic screens such as the ``ideal'' checkerboard geometry shown in Fig.~\ref{fig:res_check}(a).
Babinet's principle ensures that the power reflectance $R$ of the original problem is equal to the power transmittance $T\sub{c}$ of the dual problem~($T\sub{c}=R$).
On the other hand, the power transmittance $T$ of the original problem must be equal to $T\sub{c}$, due to the self-duality~($T=T\sub{c}$).
Then, the energy conservation law~($T+R=T+T\sub{c}=1$) gives the result that the power transmittance of electromagnetic waves
through the checkerboard screen is equal to $1/2$ and is independent of the frequency of the incident waves~\cite{Kempa2006}.
However, this result seems strange for the following reasons: (i)~the ideal checkerboard structure shows a frequency-independent response in spite of the fact that its geometry has periodicity or a characteristic length;
(ii)~the frequency-independent spectrum violates Foster's reactance theorem in the long-wavelength limit, which states that the reactance of passive and lossless systems must strictly increase with frequency~\cite{Chen2013}.

During the past few decades, several experimental and numerical attempts have been made to observe the unusual frequency-independent transmission spectrum of the ideal checkerboard structures~\cite{compton1984babinet,Singh:09,Takano2014}, but none of them has succeeded in observing the flat spectrum predicted by this theory.
This inconsistency is attributed to the point contacts at the corners of the metal patches in the ideal checkerboard geometry illustrated in Fig.~\ref{fig:res_check}(b).
In principle, it is impossible to realize such ideal point contacts.
Thus, the corners of actual structures have to be either connected or disconnected.
It is known that such metallic objects that are nearly touching exhibit singular electromagnetic responses~\cite{Pendry2013}.
In addition, as is the case of the dc electrical conduction of the system~\cite{Sheng1982}, the connectivity of the metal corners has a significant influence over the scattering characteristics at higher frequencies~\cite{Takano2014,Edmunds2010,Ramakrishna2011}.
This critical behavior has been explained by percolation theory with the identification of the ideal checkerboard as the structure representing the percolation threshold between connected structures and disconnected ones~\cite{Kempa2006}.
In the case of self-complementary antennas, the connectivity of the metal is not considered as a critical problem, because the point contacts work as feed points to connect external circuits.
Therefore, the existence of a flat transmission spectrum in metallic checkerboard screens is still controversial and worth pursuing.

In this Letter, to resolve the long-standing problem concerning self-complementarity, we propose using yet another intermediate state between connection and disconnection instead of the singular point contacts.
To be more specific, we will replace the corners of the metallic checkerboard structure with resistive sheets, as shown in Fig.~\ref{fig:res_check}(c).
By controlling the resistance of the sheets, we can realize the intermediates between the connected states and the disconnected ones \footnote{Note that in circuit theory an infinite resistance corresponds to an open circuit, and zero resistance corresponds to a short circuit~\cite{Desoer1969basic}.}.
We then introduce an extension of Babinet's principle to deal with the resistive elements and show a self-dual condition on the resistance. Finally, we experimentally demonstrate in the terahertz regime that the checkerboard structure loaded with resistive elements exhibits the predicted frequency-independent response under this specific condition on the resistance.

%
%
\begin{figure}[!b]
 \includegraphics[width=7.8cm]{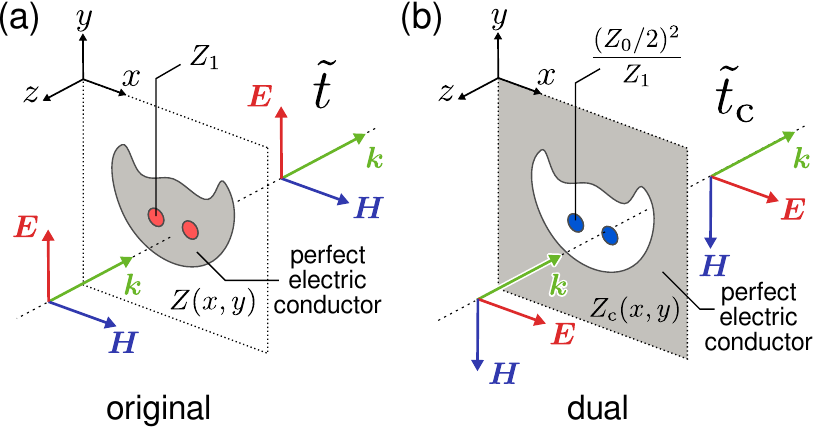}
 \caption{\label{fig:babinet}(color online). Two problems of plane-wave scattering related through Babinet's principle extended to finite sheet impedance. (a)~The original problem. (b)~The dual problem with the complementary sheet-impedance distribution and the incident polarization orthogonal to the original one. The symbols $\vct{E}$, $\vct{H}$, and $\vct{k}$ represent the electric field, the magnetic field, and the wave vector, respectively.}
\end{figure}
The familiar form of Babinet's principle relates the scattering due to a thin metallic structure to that due to its complement~\cite{JacksonBook}. Note that the structures are assumed to be perfect electric conductors.
To discuss the effect of the resistive elements in the resistive checkerboard structure, we need to extend Babinet's principle to finite sheet impedance.
Let us consider the problem of plane-wave scattering by a structure placed in a vacuum for which the spatial distribution of the sheet impedance $Z(x,y)$ is as shown in Fig.~\ref{fig:babinet}(a), where $(x,y)$ gives the coordinates on the structure.
Next, we construct the dual problem, shown in Fig.~\ref{fig:babinet}(b), where the complementary sheet-impedance distribution $Z\sub{c}(x,y)$ is defined as
\begin{equation}
 Z\sub{c}(x,y) = \frac{(Z_0/2)^2}{Z(x,y)}.\label{eq:inversion}
\end{equation}
Here, $Z_0\sim 377\,\U{O}$ is the impedance of a vacuum.
In simple terms, the sheet impedance is inverted at each point on the structure.
We note that the polarizations of the incident waves of the two problems differ by 90 deg.
Babinet's principle for finite sheet impedance relates the transmittance of these two situations~\cite{BaumNote1974,Nakata2013}:
\begin{equation}
 \tilde{t}(\omega) + \tilde{t}\sub{c}(\omega)=1,\label{eq:Babinet}
\end{equation}
where $\tilde{t}(\omega)$ and $\tilde{t}\sub{c}(\omega)$ are the complex amplitude transmittance of the zeroth-order diffraction mode in the original problem and its dual, respectively,  and $\omega$ is the angular frequency of the incident wave. Here, the zeroth-order diffraction mode refers to the mode that has the same wave vector and polarization as the incident one.
It is easy to confirm that the extended version of Babinet's principle includes the conventional one by recalling that $Z=0$ and $Z=\infty$ correspond to perfect electric conductors and holes, respectively.
We emphasize that Babinet's principle relates the amplitude coefficients, not the power transmittance.
In cases with neither polarization conversion nor diffraction, the power transmittance is related~($T+T\sub{c}=1$) by a combination of Babinet's principle~($T\sub{c}=R$) and energy conservation~($T+R=1$).

We apply this principle to the resistive checkerboard structure.
If the sheet impedance of the resistive sheet is equal to $Z_0/2$, the structure is self-complementary, because resistive sheets are invariant under the transformation in Eq.~(\ref{eq:inversion}) and the complementary structure can overlap the original one. 
Hence, we see that the dual problem of the resistive checkerboard structure is identical to the original one when linearly polarized plane waves are normally incident. Consequently, we obtain $\tilde{t}(\omega)=\tilde{t}\sub{c}(\omega)$, and combining this with Eq.~(\ref{eq:Babinet}) leads to
\begin{equation}
 \label{eq:half}
\tilde{t}(\omega)=\tilde{t}\sub{c}(\omega)=\frac{1}{2}.
\end{equation}
This indicates that the resistive checkerboard structure shows a frequency-independent transmission spectrum if the sheet impedance of the resistive sheets is equal to $Z_0/2$.
The same applies to the reflection spectrum, and we obtain an amplitude reflectance~$\tilde{r}(\omega)=-1/2$.
Therefore, half the incident power must be diffracted or absorbed. In particular, in the long-wavelength limit where there is no diffraction, half the power is absorbed by the resistive sheets.
The theoretical general sufficient conditions for the frequency-independent response have been presented in our previous paper~(see Ref.~\cite{Nakata2013}).
It should be noted that $Z_0/2$ is replaced with $Z_0/(2n)$ when the structure is surrounded by an isotropic dielectric medium with refractive index $n$.
\begin{figure}[!b]
 \includegraphics[width=7.5cm]{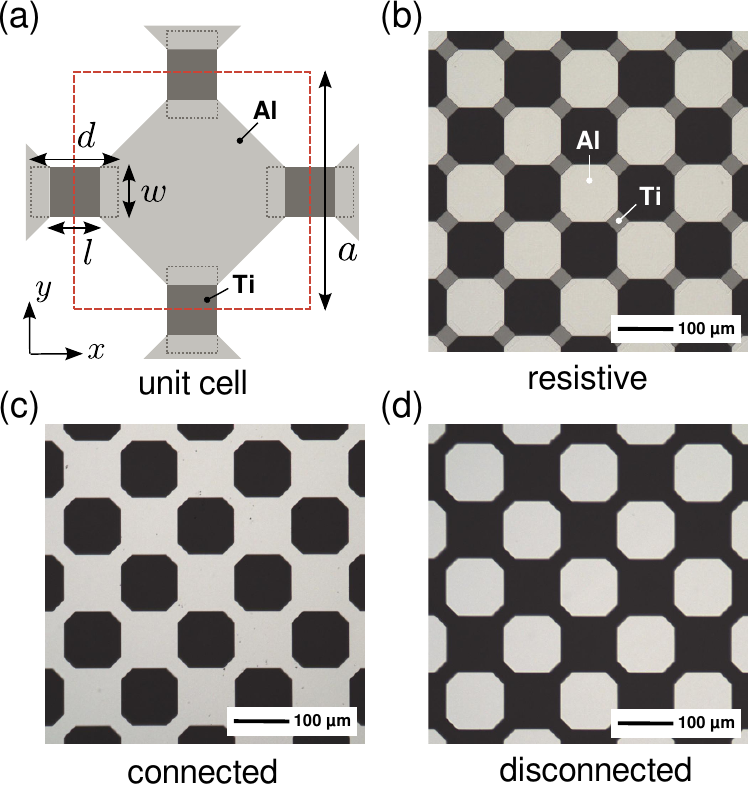}
 \caption{\label{fig:samples}(color online). (a)~The unit cell of the resistive checkerboard structure. The dimensions are as follows: $a=150\,\U{um}$, $d=50\,\U{um}$,  $w=30\,\U{um}$, and $l=30\,\U{um}$. The gray dotted lines indicate the area of overlap of the two layers. (b)--(d)~Photomicrographs of the fabricated checkerboard structures.}
\end{figure}
\begin{figure*}[t]
 \includegraphics[width=15cm]{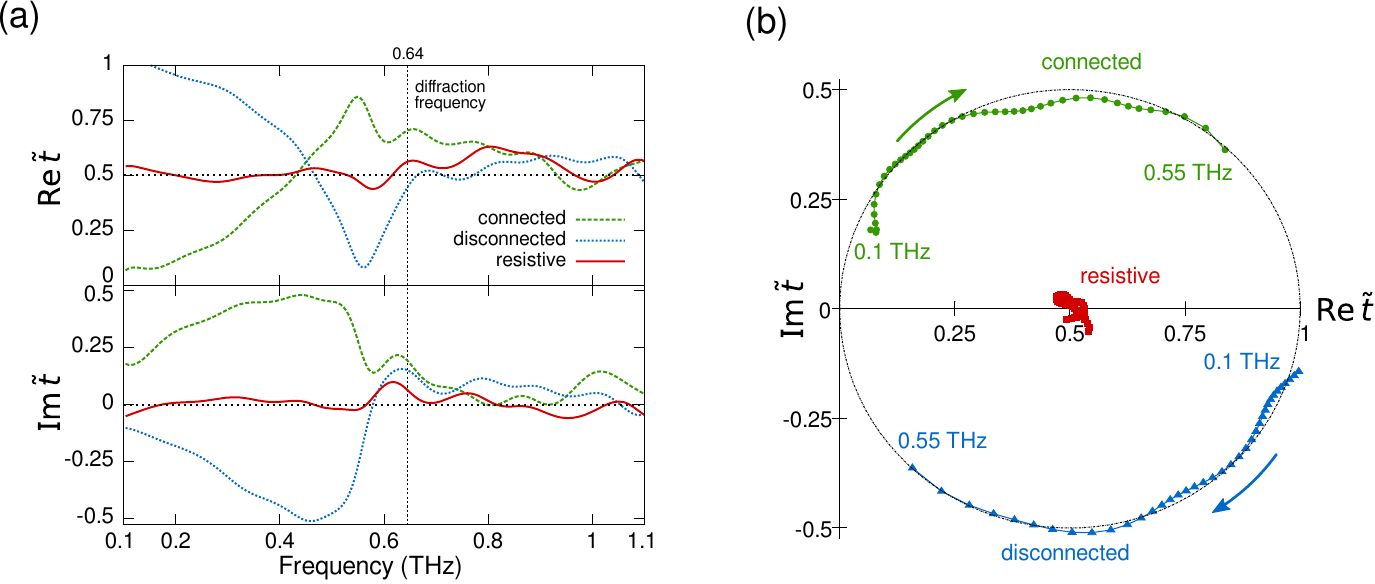}
 \caption{\label{fig:results}(color online). The results of the transmission measurements. (a)~The transmission spectra. (b)~The loci of the amplitude transmission coefficients in the frequency range of $0.1$--$0.55\,\U{THz}$.}
\end{figure*}

%
%
We will now explain our experimental demonstration of the above situation. 
The resistive checkerboard structure was fabricated on a $c$-cut sapphire substrate~($20\,\U{mm}\times 20\,\U{mm} \times 900\,\U{um}$, Kyocera) by using the standard photolithography and lift-off technique.
 For the case of normal incidence, the refractive index of the $c$-cut sapphire plate is that for ordinary waves, i.e., $n\sub{Sa}\sim 3.1$ in the terahertz regime~\cite{Grischkowsky1990}. 
The unit cell of the fabricated resistive checkerboard structure is shown in Fig.~\ref{fig:samples}(a), and its photomicrograph is shown in Fig.~\ref{fig:samples}(b).
The structure consists of two layers. A resistive titanium~(Ti) layer 19~$\U{nm}$ thick and a conductive aluminum~(Al) layer 400~$\U{nm}$ thick were deposited by electron-beam evaporation at room temperature.
The Ti rectangles have a margin of $d-l=20\,\U{um}$ that overlaps with the Al patches to assure electrical contact.
With the terahertz time-domain spectroscopy~(THz-TDS) technique~(see the Supplemental Material for the experimental details~\footnote{See Supplemental Material at [URL], which includes Ref.~\cite{Crooker2002}, for the experimental details and additional numerical simulations.})\nocite{Crooker2002} and Tinkham's equation~\cite{Walther2007}, the sheet impedance of the 19-$\U{nm}$-thick Ti film was estimated to be $0.98\times Z_0/(2n\sub{Sa})$.
The thickness of the Al layer was determined by taking into account the skin depth of terahertz radiation in Al, $\sim$$100\,\U{nm}$ at $1\,\U{THz}$~\cite{Azad2005}.
For comparison, the connected and disconnected checkerboard structures were also fabricated as shown in Figs.~\ref{fig:samples}(c) and \ref{fig:samples}(d). Their dimensions are the same as the resistive one, although they comprise only an Al layer and do not have self-complementarity.

To investigate their transmission properties, the fabricated structures were also characterized with THz-TDS~\cite{Note2}.
The probe terahertz beam was linearly polarized in the $y$~direction and focused on the samples under normal incidence with respect to the sample plane.
In order to apply Babinet's principle, it is necessary to satisfy the mirror-symmetry condition with respect to the sample plane. During the measurements, the surfaces of the checkerboard structures were covered with another plain sapphire plate so that the structures were symmetrically sandwiched between sapphire plates.
We used a pair of plain sapphire plates as a reference.
The amplitude transmission coefficients $\tilde{t}(\omega)$ of the samples were calculated by $\tilde{t}(\omega)=\tilde{E}\sub{sample}(\omega)/\tilde{E}\sub{sapphire}(\omega)$, where $\tilde{E}\sub{sample\,(sapphire)}$ represents the Fourier transform of a recorded electric field of a terahertz pulse transmitted through a sample~(sapphire).
In the calculation, the echo pulses caused by reflections at the boundaries of the substrate were removed from the temporal waveforms by multiplying by a time window.
We note that the effect of the diffracted waves is negligible if the detector is sufficiently distant from the samples, and the terahertz beam can be approximated by a plane wave in the neighborhood of the focal point.
Therefore, we can regard the experimental results as the theoretical transmission coefficients of the zeroth-order diffraction mode.

%
%
The measured amplitude transmission spectra of the checkerboard structures are shown in Fig.~\ref{fig:results}(a)~(see Fig.~S2 in the Supplemental Material for comparison with numerical simulations~\cite{Note2}).
It is clearly confirmed from the result that the resistive case shows a nearly frequency-independent spectrum in both the real and imaginary parts.
Note that the flatness continues beyond the diffraction frequency~($0.64\,\U{THz}$) or the homogenization limit.
On the other hand, the spectra of the connected and disconnected checkerboard structures highly depend on the frequencies of the incident waves.
Figure~\ref{fig:results}(b) shows the loci of the amplitude transmission coefficients in the frequency range of $0.1$--$0.55\,\U{THz}$, where diffraction is negligible.
We can easily confirm that the locus of the resistive case stays in close proximity to $(\Re \tilde{t},\Im \tilde{t})=(0.5,0)$. 
On the other hand, as the frequency increases, the loci of the connected and disconnected cases move in a clockwise direction along a circle centered at $(0.5,0)$ with radius $0.5$.
The constraint of the motion to the path defined by this circle is due to energy conservation in the nondiffraction regime~\cite{Ulrich1967}.

%
%
\begin{figure}[!t]
 \includegraphics[width=7.5cm]{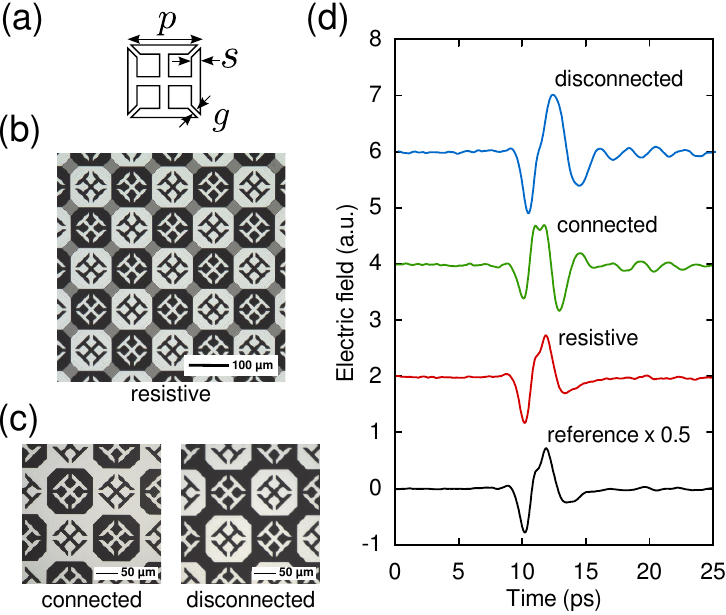}
 \caption{\label{fig:ELC_trans}(color online). (a)~The dimensions of the ELC resonators and the complementary ones, which are built in resistive checkerboard structures: $p=65\,\U{um}$, $s=8\,\U{um}$, and $g=8\,\U{um}$. (b)~A photomicrograph of the resistive self-complementary metamaterial. (c)~Photomicrographs of the connected and disconnected checkerboard metamaterials. (d)~The recorded temporal waveforms of the terahertz electric fields that were transmitted through a sapphire reference and the disconnected, connected, and resistive metamaterials. The curves are vertically offset by 2 units for the sake of clarity.}
\end{figure}
Equation~(\ref{eq:half}) holds for any self-complementarity structures with $n$-fold rotational symmetry~($n\geq 3$)~\cite{Nakata2013}. The consequent frequency-independent response is counterintuitive, especially when these are resonant structures, which are often building blocks of metamaterials. Here, we consider resistive self-complementary structures loaded with electric-inductive-capacitive~(ELC) resonators~\cite{2006ApPhL..88d1109S}, and their complementary structures~\cite{Chen:07}.
Their shape and dimensions are shown in Fig.~\ref{fig:ELC_trans}(a).
They are designed to exhibit an inductive-capacitive resonance at $0.41\,\U{THz}$ when surrounded by sapphire.
A photomicrograph of the fabricated planar metamaterial is shown in Fig.~\ref{fig:ELC_trans}(b).
The details of the fabrication are the same as above.
The ELC resonators and their complements are placed at the centers of the metallic patches or holes in the resistive checkerboard structure to maintain self-complementarity.
The connected and disconnected structures were also prepared for comparison, as shown in Fig.~\ref{fig:ELC_trans}(c).
We measured the electric fields transmitted through the samples by THz-TDS.
Figure~\ref{fig:ELC_trans}(d) shows the measured waveforms of the electric fields after passing through the reference sapphire plates and the metamaterials~(see Fig.~S7 in the Supplemental Material for their transmission spectra~\cite{Note2}).
In both the connected and the disconnected cases, the waveforms are highly distorted as compared to those of the reference; this is due to the frequency-dependent transmission characteristics of the metamaterials.
In addition, we can see persistent oscillations after the main pulse, and these are the evidence that the incident energy is temporarily stored in the resonant structures and then released afterward.
On the other hand, the distortion in the resistive case is obviously small, and persistent oscillations are not observed.
This clearly indicates that self-complementarity suppresses the frequency dependence of the transmission induced by the resonant structures.
We note that resistive loading without self-complementarity does not cause the frequency independence~(see Fig.~S5 in the Supplemental Material~\cite{Note2}).

%
%
In conclusion, we performed an experiment that demonstrated the controversial frequency-independent transmission spectra of self-complementary screens by replacing singular point contacts with resistive sheets.
Consequently, this work revealed that the controversy arose from an implicit assumption about the power transmittance,~$T+T\sub{c}=1$, which does not hold in self-dual cases.
We also showed that self-dual symmetry can suppress even the strongly frequency-dependent response of the resonant structure.
The phenomenon reported here is universal, because it is based on the self-duality of the system.
Thus, it is expected to play an important role in a broad range of the electromagnetic spectrum and to be helpful for practical applications such as designing broadband metamaterials.
In addition, the flat transmission and reflection characteristics of the self-complementary screens are applicable to broadband coherent perfect absorption in artificial structures~\cite{Nakata2013,Pu2012}.
We note that there is no theoretical limitation on the size of the resistive sheets and that the energy of the incoming radiation can be concentrated on highly subwavelength regions; thus this has the potential to enhance nonlinear phenomena and light-harvesting applications.
Finally, we point out that similar nearly flat transmission spectra in finite frequency ranges have been observed in thin metal films close to the metal-insulator transition, where metal islands are randomly connected by lossy narrow necks~\cite{Yagil1987,Davis1991}.
Although these involve imperfections such as randomness, broken mirror symmetry, and material dispersion, they can be regarded as statistically and approximately self-dual screens. Thus, the perspective of this Letter provides another physical insight into their behavior, beyond the conventional view from effective medium theory~\cite{Sarychev1995}.

%
%
\begin{acknowledgments}
This study was deeply inspired by ideas of M.~Hangyo of Osaka University, who passed away recently.
The authors gratefully acknowledge fruitful discussions with K.~Takano and Y.~Tanaka and their experimental support. We thank F.~Miyamaru for his technical advice in the THz-TDS experiments. We are also grateful to R.\,C.~McPhedran for a useful comment at a conference held in Bordeaux and to H.\,Nakano for giving us information on a research report. 
This work was supported in part by JSPS KAKENHI Grants No.~22109004, No.~25790065, and No.~25287101. The samples were prepared with the help of Kyoto University Nano Technology Hub in the ``Nanotechnology Platform Project'' sponsored by MEXT of Japan. Y.\,U. and Y.\,N. were supported by JSPS Research Fellowships for Young Scientists.
\end{acknowledgments}
%
%

\begin{thebibliography}{36}%
\makeatletter
\providecommand \@ifxundefined [1]{%
 \@ifx{#1\undefined}
}%
\providecommand \@ifnum [1]{%
 \ifnum #1\expandafter \@firstoftwo
 \else \expandafter \@secondoftwo
 \fi
}%
\providecommand \@ifx [1]{%
 \ifx #1\expandafter \@firstoftwo
 \else \expandafter \@secondoftwo
 \fi
}%
\providecommand \natexlab [1]{#1}%
\providecommand \enquote  [1]{``#1''}%
\providecommand \bibnamefont  [1]{#1}%
\providecommand \bibfnamefont [1]{#1}%
\providecommand \citenamefont [1]{#1}%
\providecommand \href@noop [0]{\@secondoftwo}%
\providecommand \href [0]{\begingroup \@sanitize@url \@href}%
\providecommand \@href[1]{\@@startlink{#1}\@@href}%
\providecommand \@@href[1]{\endgroup#1\@@endlink}%
\providecommand \@sanitize@url [0]{\catcode `\\12\catcode `\$12\catcode
  `\&12\catcode `\#12\catcode `\^12\catcode `\_12\catcode `\%12\relax}%
\providecommand \@@startlink[1]{}%
\providecommand \@@endlink[0]{}%
\providecommand \url  [0]{\begingroup\@sanitize@url \@url }%
\providecommand \@url [1]{\endgroup\@href {#1}{\urlprefix }}%
\providecommand \urlprefix  [0]{URL }%
\providecommand \Eprint [0]{\href }%
\providecommand \doibase [0]{http://dx.doi.org/}%
\providecommand \selectlanguage [0]{\@gobble}%
\providecommand \bibinfo  [0]{\@secondoftwo}%
\providecommand \bibfield  [0]{\@secondoftwo}%
\providecommand \translation [1]{[#1]}%
\providecommand \BibitemOpen [0]{}%
\providecommand \bibitemStop [0]{}%
\providecommand \bibitemNoStop [0]{.\EOS\space}%
\providecommand \EOS [0]{\spacefactor3000\relax}%
\providecommand \BibitemShut  [1]{\csname bibitem#1\endcsname}%
\let\auto@bib@innerbib\@empty
\bibitem [{\citenamefont {Bliokh}\ \emph {et~al.}(2013)\citenamefont {Bliokh},
  \citenamefont {Bekshaev},\ and\ \citenamefont {Nori}}]{Bliokh2013}%
  \BibitemOpen
  \bibfield  {author} {\bibinfo {author} {\bibfnamefont {K.~Y.}\ \bibnamefont
  {Bliokh}}, \bibinfo {author} {\bibfnamefont {A.~Y.}\ \bibnamefont
  {Bekshaev}}, \ and\ \bibinfo {author} {\bibfnamefont {F.}~\bibnamefont
  {Nori}},\ }\href {\doibase 10.1088/1367-2630/15/3/033026} {\bibfield
  {journal} {\bibinfo  {journal} {New J. Phys.}\ }\textbf {\bibinfo {volume}
  {15}},\ \bibinfo {pages} {033026} (\bibinfo {year} {2013})}\ \BibitemShut
  {NoStop}%
\bibitem [{\citenamefont {Alvarez}\ \emph {et~al.}(1995)\citenamefont
  {Alvarez}, \citenamefont {Alvarez-Gaume},\ and\ \citenamefont
  {Lozano}}]{Alvarez1995}%
  \BibitemOpen
  \bibfield  {author} {\bibinfo {author} {\bibfnamefont {E.}~\bibnamefont
  {Alvarez}}, \bibinfo {author} {\bibfnamefont {L.}~\bibnamefont
  {Alvarez-Gaume}}, \ and\ \bibinfo {author} {\bibfnamefont {Y.}~\bibnamefont
  {Lozano}},\ }\href
  {http://www.sciencedirect.com/science/article/pii/092056329500429D}
  {\bibfield  {journal} {\bibinfo  {journal} {Nucl. Phys. B (Proc. Suppl.)}\
  }\textbf {\bibinfo {volume} {41}},\ \bibinfo {pages} {1} (\bibinfo {year}
  {1995})}\ \BibitemShut {NoStop}%
\bibitem [{\citenamefont {Desoer}\ and\ \citenamefont
  {Kuh}(1969)}]{Desoer1969basic}%
  \BibitemOpen
  \bibfield  {author} {\bibinfo {author} {\bibfnamefont {C.}~\bibnamefont
  {Desoer}}\ and\ \bibinfo {author} {\bibfnamefont {E.}~\bibnamefont {Kuh}},\
  }\href@noop {} {\emph {\bibinfo {title} {Basic Circuit Theory}}}\ (\bibinfo
  {publisher} {McGraw-Hill Kogakusha},\ \bibinfo {address} {Tokyo},\ \bibinfo
  {year} {1969})\BibitemShut {NoStop}%
\bibitem [{\citenamefont {Kramers}\ and\ \citenamefont
  {Wannier}(1941)}]{Kramers1941}%
  \BibitemOpen
  \bibfield  {author} {\bibinfo {author} {\bibfnamefont {H.~A.}\ \bibnamefont
  {Kramers}}\ and\ \bibinfo {author} {\bibfnamefont {G.~H.}\ \bibnamefont
  {Wannier}},\ }\href@noop {} {\bibfield  {journal} {\bibinfo  {journal} {Phys.
  Rev.}\ }\textbf {\bibinfo {volume} {60}},\ \bibinfo {pages} {252} (\bibinfo
  {year} {1941})}\ \BibitemShut {NoStop}%
\bibitem [{\citenamefont {Keller}(1964)}]{Keller1964}%
  \BibitemOpen
  \bibfield  {author} {\bibinfo {author} {\bibfnamefont {J.~B.}\ \bibnamefont
  {Keller}},\ }\href {\doibase 10.1063/1.1704146} {\bibfield  {journal}
  {\bibinfo  {journal} {J. Math. Phys.}\ }\textbf {\bibinfo {volume} {5}},\
  \bibinfo {pages} {548} (\bibinfo {year} {1964})}\ \BibitemShut {NoStop}%
\bibitem [{\citenamefont {Dykhne}(1970)}]{Dykhne1970}%
  \BibitemOpen
  \bibfield  {author} {\bibinfo {author} {\bibfnamefont {A.~M.}\ \bibnamefont
  {Dykhne}},\ }\href@noop {} {\bibfield  {journal} {\bibinfo  {journal} {Zh.
  Eksp. Teor. Fiz.}\ }\textbf {\bibinfo {volume} {59}},\ \bibinfo {pages} {110}
  (\bibinfo {year} {1970})}\ \translation{Sov. Phys. JETP \textbf{32}, 63
  (1971)}\BibitemShut {NoStop}%
\bibitem [{\citenamefont {Lin}(1967)}]{LinSelfDual}%
  \BibitemOpen
  \bibfield  {author} {\bibinfo {author} {\bibfnamefont {P.~M.}\ \bibnamefont
  {Lin}},\ }\href@noop {} {\bibfield  {journal} {\bibinfo  {journal} {{IEEE}
  Trans. Circuit Theory}\ }\textbf {\bibinfo {volume} {14}},\ \bibinfo {pages}
  {172} (\bibinfo {year} {1967})}\ \BibitemShut {NoStop}%
\bibitem [{\citenamefont {Mushiake}(1996)}]{Mushiakebook}%
  \BibitemOpen
  \bibfield  {author} {\bibinfo {author} {\bibfnamefont {Y.}~\bibnamefont
  {Mushiake}},\ }\href@noop {} {\emph {\bibinfo {title} {{Self-Complementary
  Antennas: Principle of Self-Complementarity for Constant Impedance}}}}\
  (\bibinfo  {publisher} {Springer},\ \bibinfo {address} {London},\ \bibinfo
  {year} {1996})\BibitemShut {NoStop}%
\bibitem [{\citenamefont {Lindell}\ \emph {et~al.}(2009)\citenamefont
  {Lindell}, \citenamefont {Sihvola}, \citenamefont {Yla-Oijala},\ and\
  \citenamefont {Wallen}}]{Lindell2009}%
  \BibitemOpen
  \bibfield  {author} {\bibinfo {author} {\bibfnamefont {I.}~\bibnamefont
  {Lindell}}, \bibinfo {author} {\bibfnamefont {A.}~\bibnamefont {Sihvola}},
  \bibinfo {author} {\bibfnamefont {P.}~\bibnamefont {Yla-Oijala}}, \ and\
  \bibinfo {author} {\bibfnamefont {H.}~\bibnamefont {Wallen}},\ }\href
  {\doibase 10.1109/TAP.2009.2027180} {\bibfield  {journal} {\bibinfo
  {journal} {IEEE Trans. Antennas Propag.}\ }\textbf {\bibinfo {volume} {57}},\
  \bibinfo {pages} {2725} (\bibinfo {year} {2009})}\ \BibitemShut {NoStop}%
\bibitem [{\citenamefont {Senechal}(1988)}]{Senechal1988}%
  \BibitemOpen
  \bibfield  {author} {\bibinfo {author} {\bibfnamefont {M.}~\bibnamefont
  {Senechal}},\ }\href {\doibase 10.1016/0898-1221(88)90244-1} {\bibfield
  {journal} {\bibinfo  {journal} {Comput. Math. Appl.}\ }\textbf {\bibinfo
  {volume} {16}},\ \bibinfo {pages} {545} (\bibinfo {year} {1988})}\
  \BibitemShut {NoStop}%
\bibitem [{\citenamefont {Jackson}(1998)}]{JacksonBook}%
  \BibitemOpen
  \bibfield  {author} {\bibinfo {author} {\bibfnamefont {J.~D.}\ \bibnamefont
  {Jackson}},\ }\href@noop {} {\emph {\bibinfo {title} {Classical
  Electrodynamics}}},\ \bibinfo {edition} {3rd}\ ed.\ (\bibinfo  {publisher}
  {Wiley},\ \bibinfo {address} {New York},\ \bibinfo {year} {1998})\BibitemShut
  {NoStop}%
\bibitem [{\citenamefont {Nakano}(2008)}]{BalanisAntenna}%
  \BibitemOpen
  \bibfield  {author} {\bibinfo {author} {\bibfnamefont {H.}~\bibnamefont
  {Nakano}},\ }in\ \href@noop {} {\emph {\bibinfo {booktitle} {Modern Antenna
  Handbook}}},\ \bibinfo {editor} {edited by\ \bibinfo {editor} {\bibfnamefont
  {C.~A.}\ \bibnamefont {Balanis}}}\ (\bibinfo  {publisher} {Wiley},\ \bibinfo
  {address} {New York},\ \bibinfo {year} {2008}),\ Chap.~\bibinfo {chapter} {6},
  pp.\ \bibinfo {pages} {263--324}\BibitemShut {NoStop}%
\bibitem [{\citenamefont {Kempa}(2010)}]{Kempa2006}%
  \BibitemOpen
  \bibfield  {author} {\bibinfo {author} {\bibfnamefont {K.}~\bibnamefont
  {Kempa}},\ }\href {\doibase 10.1002/pssr.201004266} {\bibfield  {journal}
  {\bibinfo  {journal} {Phys. Status Solidi Rapid Res. Lett.}\ }\textbf
  {\bibinfo {volume} {4}},\ \bibinfo {pages} {218} (\bibinfo {year} {2010})}\
  \BibitemShut {NoStop}%
\bibitem [{\citenamefont {Chen}\ \emph {et~al.}(2013)\citenamefont {Chen},
  \citenamefont {Argyropoulos},\ and\ \citenamefont {Al\`{u}}}]{Chen2013}%
  \BibitemOpen
  \bibfield  {author} {\bibinfo {author} {\bibfnamefont {P.-Y.}\ \bibnamefont
  {Chen}}, \bibinfo {author} {\bibfnamefont {C.}~\bibnamefont {Argyropoulos}},
  \ and\ \bibinfo {author} {\bibfnamefont {A.}~\bibnamefont {Al\`{u}}},\ }\href
  {\doibase 10.1103/PhysRevLett.111.233001} {\bibfield  {journal} {\bibinfo
  {journal} {Phys. Rev. Lett.}\ }\textbf {\bibinfo {volume} {111}},\ \bibinfo
  {pages} {233001} (\bibinfo {year} {2013})}\ \BibitemShut {NoStop}%
\bibitem [{\citenamefont {Compton}\ \emph {et~al.}(1984)\citenamefont
  {Compton}, \citenamefont {Macfarlane}, \citenamefont {Whitbourn},
  \citenamefont {Blanco},\ and\ \citenamefont
  {McPhedran}}]{compton1984babinet}%
  \BibitemOpen
  \bibfield  {author} {\bibinfo {author} {\bibfnamefont {R.~C.}\ \bibnamefont
  {Compton}}, \bibinfo {author} {\bibfnamefont {J.~C.}\ \bibnamefont
  {Macfarlane}}, \bibinfo {author} {\bibfnamefont {L.~B.}\ \bibnamefont
  {Whitbourn}}, \bibinfo {author} {\bibfnamefont {M.~M.}\ \bibnamefont
  {Blanco}}, \ and\ \bibinfo {author} {\bibfnamefont {R.~C.}\ \bibnamefont
  {McPhedran}},\ }\href@noop {} {\bibfield  {journal} {\bibinfo  {journal}
  {Opt. Acta}\ }\textbf {\bibinfo {volume} {31}},\ \bibinfo {pages} {515}
  (\bibinfo {year} {1984})}\ \BibitemShut {NoStop}%
\bibitem [{\citenamefont {Singh}\ \emph {et~al.}(2009)\citenamefont {Singh},
  \citenamefont {Rockstuhl}, \citenamefont {Menzel}, \citenamefont {Meyrath},
  \citenamefont {He}, \citenamefont {Giessen}, \citenamefont {Lederer},\ and\
  \citenamefont {Zhang}}]{Singh:09}%
  \BibitemOpen
  \bibfield  {author} {\bibinfo {author} {\bibfnamefont {R.}~\bibnamefont
  {Singh}}, \bibinfo {author} {\bibfnamefont {C.}~\bibnamefont {Rockstuhl}},
  \bibinfo {author} {\bibfnamefont {C.}~\bibnamefont {Menzel}}, \bibinfo
  {author} {\bibfnamefont {T.~P.}\ \bibnamefont {Meyrath}}, \bibinfo {author}
  {\bibfnamefont {M.}~\bibnamefont {He}}, \bibinfo {author} {\bibfnamefont
  {H.}~\bibnamefont {Giessen}}, \bibinfo {author} {\bibfnamefont
  {F.}~\bibnamefont {Lederer}}, \ and\ \bibinfo {author} {\bibfnamefont
  {W.}~\bibnamefont {Zhang}},\ }\href {\doibase 10.1364/OE.17.009971}
  {\bibfield  {journal} {\bibinfo  {journal} {Opt. Express}\ }\textbf {\bibinfo
  {volume} {17}},\ \bibinfo {pages} {9971} (\bibinfo {year} {2009})}\
  \BibitemShut {NoStop}%
\bibitem [{\citenamefont {Takano}\ \emph {et~al.}(2014)\citenamefont {Takano},
  \citenamefont {Miyamaru}, \citenamefont {Akiyama}, \citenamefont {Miyazaki},
  \citenamefont {Takeda}, \citenamefont {Abe}, \citenamefont {Tokuda},
  \citenamefont {Ito},\ and\ \citenamefont {Hangyo}}]{Takano2014}%
  \BibitemOpen
  \bibfield  {author} {\bibinfo {author} {\bibfnamefont {K.}~\bibnamefont
  {Takano}}, \bibinfo {author} {\bibfnamefont {F.}~\bibnamefont {Miyamaru}},
  \bibinfo {author} {\bibfnamefont {K.}~\bibnamefont {Akiyama}}, \bibinfo
  {author} {\bibfnamefont {H.}~\bibnamefont {Miyazaki}}, \bibinfo {author}
  {\bibfnamefont {M.~W.}\ \bibnamefont {Takeda}}, \bibinfo {author}
  {\bibfnamefont {Y.}~\bibnamefont {Abe}}, \bibinfo {author} {\bibfnamefont
  {Y.}~\bibnamefont {Tokuda}}, \bibinfo {author} {\bibfnamefont
  {H.}~\bibnamefont {Ito}}, \ and\ \bibinfo {author} {\bibfnamefont
  {M.}~\bibnamefont {Hangyo}},\ }\href {\doibase 10.1364/OE.22.024787}
  {\bibfield  {journal} {\bibinfo  {journal} {Opt. Express}\ }\textbf {\bibinfo
  {volume} {22}},\ \bibinfo {pages} {24787} (\bibinfo {year} {2014})}\
  \BibitemShut {NoStop}%
\bibitem [{\citenamefont {Pendry}\ \emph {et~al.}(2013)\citenamefont {Pendry},
  \citenamefont {Fern\'{a}ndez-Dom\'{\i}nguez}, \citenamefont {Luo},\ and\
  \citenamefont {Zhao}}]{Pendry2013}%
  \BibitemOpen
  \bibfield  {author} {\bibinfo {author} {\bibfnamefont {J.~B.}\ \bibnamefont
  {Pendry}}, \bibinfo {author} {\bibfnamefont {A.~I.}\ \bibnamefont
  {Fern\'{a}ndez-Dom\'{\i}nguez}}, \bibinfo {author} {\bibfnamefont
  {Y.}~\bibnamefont {Luo}}, \ and\ \bibinfo {author} {\bibfnamefont
  {R.}~\bibnamefont {Zhao}},\ }\href {\doibase 10.1038/nphys2667} {\bibfield
  {journal} {\bibinfo  {journal} {Nat. Phys.}\ }\textbf {\bibinfo {volume}
  {9}},\ \bibinfo {pages} {518} (\bibinfo {year} {2013})}\ \BibitemShut
  {NoStop}%
\bibitem [{\citenamefont {Sheng}\ and\ \citenamefont {Kohn}(1982)}]{Sheng1982}%
  \BibitemOpen
  \bibfield  {author} {\bibinfo {author} {\bibfnamefont {P.}~\bibnamefont
  {Sheng}}\ and\ \bibinfo {author} {\bibfnamefont {R.~V.}\ \bibnamefont
  {Kohn}},\ }\href
  {http://journals.aps.org/prb/abstract/10.1103/PhysRevB.26.1331} {\bibfield
  {journal} {\bibinfo  {journal} {Phys. Rev. B}\ }\textbf {\bibinfo {volume}
  {26}},\ \bibinfo {pages} {1331} (\bibinfo {year} {1982})}\ \BibitemShut
  {NoStop}%
\bibitem [{\citenamefont {Edmunds}\ \emph {et~al.}(2010)\citenamefont
  {Edmunds}, \citenamefont {Hibbins}, \citenamefont {Sambles},\ and\
  \citenamefont {Youngs}}]{Edmunds2010}%
  \BibitemOpen
  \bibfield  {author} {\bibinfo {author} {\bibfnamefont {J.~D.}\ \bibnamefont
  {Edmunds}}, \bibinfo {author} {\bibfnamefont {A.~P.}\ \bibnamefont
  {Hibbins}}, \bibinfo {author} {\bibfnamefont {J.~R.}\ \bibnamefont
  {Sambles}}, \ and\ \bibinfo {author} {\bibfnamefont {I.~J.}\ \bibnamefont
  {Youngs}},\ }\href {\doibase 10.1088/1367-2630/12/6/063007} {\bibfield
  {journal} {\bibinfo  {journal} {New J. Phys.}\ }\textbf {\bibinfo {volume}
  {12}},\ \bibinfo {pages} {063007} (\bibinfo {year} {2010})}\ \BibitemShut
  {NoStop}%
\bibitem [{\citenamefont {Ramakrishna}\ \emph {et~al.}(2011)\citenamefont
  {Ramakrishna}, \citenamefont {Mandal}, \citenamefont {Jeyadheepan},
  \citenamefont {Shukla}, \citenamefont {Chakrabarti}, \citenamefont {Kadic},
  \citenamefont {Enoch},\ and\ \citenamefont {Guenneau}}]{Ramakrishna2011}%
  \BibitemOpen
  \bibfield  {author} {\bibinfo {author} {\bibfnamefont {S.~A.}\ \bibnamefont
  {Ramakrishna}}, \bibinfo {author} {\bibfnamefont {P.}~\bibnamefont {Mandal}},
  \bibinfo {author} {\bibfnamefont {K.}~\bibnamefont {Jeyadheepan}}, \bibinfo
  {author} {\bibfnamefont {N.}~\bibnamefont {Shukla}}, \bibinfo {author}
  {\bibfnamefont {S.}~\bibnamefont {Chakrabarti}}, \bibinfo {author}
  {\bibfnamefont {M.}~\bibnamefont {Kadic}}, \bibinfo {author} {\bibfnamefont
  {S.}~\bibnamefont {Enoch}}, \ and\ \bibinfo {author} {\bibfnamefont
  {S.}~\bibnamefont {Guenneau}},\ }\href {\doibase 10.1103/PhysRevB.84.245424}
  {\bibfield  {journal} {\bibinfo  {journal} {Phys. Rev. B}\ }\textbf {\bibinfo
  {volume} {84}},\ \bibinfo {pages} {245424} (\bibinfo {year} {2011})}\
  \BibitemShut {NoStop}%
\bibitem [{Note1()}]{Note1}%
  \BibitemOpen
  \bibinfo {note} {Note that in circuit theory an infinite resistance
  corresponds to an open circuit, and zero resistance corresponds to a short
  circuit~\cite {Desoer1969basic}.}\BibitemShut {Stop}%
\bibitem [{\citenamefont {Baum}\ and\ \citenamefont
  {Singaraju}(1974)}]{BaumNote1974}%
  \BibitemOpen
  \bibfield  {author} {\bibinfo {author} {\bibfnamefont {C.~E.}\ \bibnamefont
  {Baum}}\ and\ \bibinfo {author} {\bibfnamefont {B.~K.}\ \bibnamefont
  {Singaraju}},\ }\href@noop {} {\bibfield  {journal} {\bibinfo  {journal}
  {Interaction Note No.~217, Air Force Weapons Laboratory, Kirtland Air Force Base, NM
  87117}}, \bibinfo {year} {1974}}\ \BibitemShut {NoStop}%
\bibitem [{\citenamefont {Nakata}\ \emph {et~al.}(2013)\citenamefont {Nakata},
  \citenamefont {Urade}, \citenamefont {Nakanishi},\ and\ \citenamefont
  {Kitano}}]{Nakata2013}%
  \BibitemOpen
  \bibfield  {author} {\bibinfo {author} {\bibfnamefont {Y.}~\bibnamefont
  {Nakata}}, \bibinfo {author} {\bibfnamefont {Y.}~\bibnamefont {Urade}},
  \bibinfo {author} {\bibfnamefont {T.}~\bibnamefont {Nakanishi}}, \ and\
  \bibinfo {author} {\bibfnamefont {M.}~\bibnamefont {Kitano}},\ }\href
  {\doibase 10.1103/PhysRevB.88.205138} {\bibfield  {journal} {\bibinfo
  {journal} {Phys.\ Rev. B}\ }\textbf {\bibinfo {volume} {88}},\ \bibinfo
  {pages} {205138} (\bibinfo {year} {2013})}\ \BibitemShut {NoStop}%
\bibitem [{\citenamefont {Grischkowsky}\ \emph {et~al.}(1990)\citenamefont
  {Grischkowsky}, \citenamefont {Keiding}, \citenamefont {van Exter},\ and\
  \citenamefont {Fattinger}}]{Grischkowsky1990}%
  \BibitemOpen
  \bibfield  {author} {\bibinfo {author} {\bibfnamefont {D.}~\bibnamefont
  {Grischkowsky}}, \bibinfo {author} {\bibfnamefont {S.}~\bibnamefont
  {Keiding}}, \bibinfo {author} {\bibfnamefont {M.}~\bibnamefont {van Exter}},
  \ and\ \bibinfo {author} {\bibfnamefont {{\relax Ch}.}~\bibnamefont
  {Fattinger}},\ }\href {http://utol.okstate.edu/papers/paper20.pdf} {\bibfield
   {journal} {\bibinfo  {journal} {J. Opt. Soc. Am. B}\ }\textbf {\bibinfo
  {volume} {7}},\ \bibinfo {pages} {2006} (\bibinfo {year} {1990})}\
  \BibitemShut {NoStop}%
\bibitem [{Note2()}]{Note2}%
  \BibitemOpen
  \bibinfo {note} {See Supplemental Material below, which includes
  Ref.~\cite {Crooker2002}, for the experimental details and additional
  numerical simulations.}\BibitemShut {Stop}%
\bibitem [{\citenamefont {Crooker}(2002)}]{Crooker2002}%
  \BibitemOpen
  \bibfield  {author} {\bibinfo {author} {\bibfnamefont {S.~A.}\ \bibnamefont
  {Crooker}},\ }\href {\doibase http://dx.doi.org/10.1063/1.1498904} {\bibfield
   {journal} {\bibinfo  {journal} {Rev. Sci. Instrum.}\ }\textbf {\bibinfo
  {volume} {73}},\ \bibinfo {pages} {3258} (\bibinfo {year} {2002})}\
  \BibitemShut {NoStop}%
\bibitem [{\citenamefont {Walther}\ \emph {et~al.}(2007)\citenamefont
  {Walther}, \citenamefont {Cooke}, \citenamefont {Sherstan}, \citenamefont
  {Hajar}, \citenamefont {Freeman},\ and\ \citenamefont
  {Hegmann}}]{Walther2007}%
  \BibitemOpen
  \bibfield  {author} {\bibinfo {author} {\bibfnamefont {M.}~\bibnamefont
  {Walther}}, \bibinfo {author} {\bibfnamefont {D.~G.}\ \bibnamefont {Cooke}},
  \bibinfo {author} {\bibfnamefont {C.}~\bibnamefont {Sherstan}}, \bibinfo
  {author} {\bibfnamefont {M.}~\bibnamefont {Hajar}}, \bibinfo {author}
  {\bibfnamefont {M.~R.}\ \bibnamefont {Freeman}}, \ and\ \bibinfo {author}
  {\bibfnamefont {F.~A.}\ \bibnamefont {Hegmann}},\ }\href {\doibase
  10.1103/PhysRevB.76.125408} {\bibfield  {journal} {\bibinfo  {journal} {Phys.
  Rev. B}\ }\textbf {\bibinfo {volume} {76}},\ \bibinfo {pages} {125408}
  (\bibinfo {year} {2007})}\ \BibitemShut {NoStop}%
\bibitem [{\citenamefont {Azad}\ and\ \citenamefont {Zhang}(2005)}]{Azad2005}%
  \BibitemOpen
  \bibfield  {author} {\bibinfo {author} {\bibfnamefont {A.~K.}\ \bibnamefont
  {Azad}}\ and\ \bibinfo {author} {\bibfnamefont {W.}~\bibnamefont {Zhang}},\
  }\href {http://www.opticsinfobase.org/abstract.cfm?\&id=85965} {\bibfield
  {journal} {\bibinfo  {journal} {Opt. Lett.}\ }\textbf {\bibinfo {volume}
  {30}},\ \bibinfo {pages} {2945} (\bibinfo {year} {2005})}\ \BibitemShut
  {NoStop}%
\bibitem [{\citenamefont {Ulrich}(1967)}]{Ulrich1967}%
  \BibitemOpen
  \bibfield  {author} {\bibinfo {author} {\bibfnamefont {R.}~\bibnamefont
  {Ulrich}},\ }\href
  {http://www.sciencedirect.com/science/article/pii/0020089167900280}
  {\bibfield  {journal} {\bibinfo  {journal} {Infrared Phys.}\ }\textbf
  {\bibinfo {volume} {7}},\ \bibinfo {pages} {37} (\bibinfo {year} {1967})}\
  \BibitemShut {NoStop}%
\bibitem [{\citenamefont {{Schurig}}\ \emph {et~al.}(2006)\citenamefont
  {{Schurig}}, \citenamefont {{Mock}},\ and\ \citenamefont
  {{Smith}}}]{2006ApPhL..88d1109S}%
  \BibitemOpen
  \bibfield  {author} {\bibinfo {author} {\bibfnamefont {D.}~\bibnamefont
  {{Schurig}}}, \bibinfo {author} {\bibfnamefont {J.~J.}\ \bibnamefont
  {{Mock}}}, \ and\ \bibinfo {author} {\bibfnamefont {D.~R.}\ \bibnamefont
  {{Smith}}},\ }\href {\doibase 10.1063/1.2166681} {\bibfield  {journal}
  {\bibinfo  {journal} {Appl. Phys. Lett.}\ }\textbf {\bibinfo {volume} {88}},\
  \bibinfo {eid} {041109} (\bibinfo {year} {2006})}\ \BibitemShut {NoStop}%
\bibitem [{\citenamefont {Chen}\ \emph {et~al.}(2007)\citenamefont {Chen},
  \citenamefont {O'Hara}, \citenamefont {Taylor}, \citenamefont {Averitt},
  \citenamefont {Highstrete}, \citenamefont {Lee},\ and\ \citenamefont
  {Padilla}}]{Chen:07}%
  \BibitemOpen
  \bibfield  {author} {\bibinfo {author} {\bibfnamefont {H.-T.}\ \bibnamefont
  {Chen}}, \bibinfo {author} {\bibfnamefont {J.~F.}\ \bibnamefont {O'Hara}},
  \bibinfo {author} {\bibfnamefont {A.~J.}\ \bibnamefont {Taylor}}, \bibinfo
  {author} {\bibfnamefont {R.~D.}\ \bibnamefont {Averitt}}, \bibinfo {author}
  {\bibfnamefont {C.}~\bibnamefont {Highstrete}}, \bibinfo {author}
  {\bibfnamefont {M.}~\bibnamefont {Lee}}, \ and\ \bibinfo {author}
  {\bibfnamefont {W.~J.}\ \bibnamefont {Padilla}},\ }\href {\doibase
  10.1364/OE.15.001084} {\bibfield  {journal} {\bibinfo  {journal} {Opt.
  Express}\ }\textbf {\bibinfo {volume} {15}},\ \bibinfo {pages} {1084}
  (\bibinfo {year} {2007})}\ \BibitemShut {NoStop}%
\bibitem [{\citenamefont {Pu}\ \emph {et~al.}(2012)\citenamefont {Pu},
  \citenamefont {Feng}, \citenamefont {Wang}, \citenamefont {Hu}, \citenamefont
  {Huang}, \citenamefont {Ma}, \citenamefont {Zhao}, \citenamefont {Wang},\
  and\ \citenamefont {Luo}}]{Pu2012}%
  \BibitemOpen
  \bibfield  {author} {\bibinfo {author} {\bibfnamefont {M.}~\bibnamefont
  {Pu}}, \bibinfo {author} {\bibfnamefont {Q.}~\bibnamefont {Feng}}, \bibinfo
  {author} {\bibfnamefont {M.}~\bibnamefont {Wang}}, \bibinfo {author}
  {\bibfnamefont {C.}~\bibnamefont {Hu}}, \bibinfo {author} {\bibfnamefont
  {C.}~\bibnamefont {Huang}}, \bibinfo {author} {\bibfnamefont
  {X.}~\bibnamefont {Ma}}, \bibinfo {author} {\bibfnamefont {Z.}~\bibnamefont
  {Zhao}}, \bibinfo {author} {\bibfnamefont {C.}~\bibnamefont {Wang}}, \ and\
  \bibinfo {author} {\bibfnamefont {X.}~\bibnamefont {Luo}},\ }\href
  {http://www.ncbi.nlm.nih.gov/pubmed/22330464} {\bibfield  {journal} {\bibinfo
   {journal} {Opt. Express}\ }\textbf {\bibinfo {volume} {20}},\ \bibinfo
  {pages} {2246} (\bibinfo {year} {2012})}\ \BibitemShut {NoStop}%
\bibitem [{\citenamefont {Yagil}\ and\ \citenamefont
  {Deutscher}(1987)}]{Yagil1987}%
  \BibitemOpen
  \bibfield  {author} {\bibinfo {author} {\bibfnamefont {Y.}~\bibnamefont
  {Yagil}}\ and\ \bibinfo {author} {\bibfnamefont {G.}~\bibnamefont
  {Deutscher}},\ }\href {\doibase 10.1016/0040-6090(87)90262-8} {\bibfield
  {journal} {\bibinfo  {journal} {Thin Solid Films}\ }\textbf {\bibinfo
  {volume} {152}},\ \bibinfo {pages} {465} (\bibinfo {year} {1987})}\
  \BibitemShut {NoStop}%
\bibitem [{\citenamefont {Davis}\ \emph {et~al.}(1991)\citenamefont {Davis},
  \citenamefont {McKenzie},\ and\ \citenamefont {McPhedran}}]{Davis1991}%
  \BibitemOpen
  \bibfield  {author} {\bibinfo {author} {\bibfnamefont {C.~A.}\ \bibnamefont
  {Davis}}, \bibinfo {author} {\bibfnamefont {D.~R.}\ \bibnamefont {McKenzie}},
  \ and\ \bibinfo {author} {\bibfnamefont {R.~C.}\ \bibnamefont {McPhedran}},\
  }\href {http://www.sciencedirect.com/science/article/pii/003040189190054H}
  {\bibfield  {journal} {\bibinfo  {journal} {Opt. Commun.}\ }\textbf {\bibinfo
  {volume} {85}},\ \bibinfo {pages} {70} (\bibinfo {year} {1991})}\
  \BibitemShut {NoStop}%
\bibitem [{\citenamefont {Sarychev}\ \emph {et~al.}(1995)\citenamefont
  {Sarychev}, \citenamefont {Bergman},\ and\ \citenamefont
  {Yagil}}]{Sarychev1995}%
  \BibitemOpen
  \bibfield  {author} {\bibinfo {author} {\bibfnamefont {A.~K.}\ \bibnamefont
  {Sarychev}}, \bibinfo {author} {\bibfnamefont {D.~J.}\ \bibnamefont
  {Bergman}}, \ and\ \bibinfo {author} {\bibfnamefont {Y.}~\bibnamefont
  {Yagil}},\ }\href {http://prb.aps.org/abstract/PRB/v51/i8/p5366\_1}
  {\bibfield  {journal} {\bibinfo  {journal} {Phys. Rev. B}\ }\textbf {\bibinfo
  {volume} {51}},\ \bibinfo {pages} {5366} (\bibinfo {year} {1995})}\
  \BibitemShut {NoStop}%
\end{thebibliography}
%
%
\clearpage


\section*{Supplementary material}

# Supplemental Material for
# "Frequency-independent Response of Self-complementary Checkerboard Screens"

Yoshiro Urade,[1, *] Yosuke Nakata,[2] Toshihiro Nakanishi,[1] and Masao Kitano[1, †]

[1]*Department of Electronic Science and Engineering, Kyoto University, Kyoto 615-8510, Japan*
[2]*Center for Energy and Environmental Science, Shinshu University, 4-17-1 Wakasato, Nagano 380-8553, Japan*
(Dated: May 19, 2015)

In this Supplemental Material, we present the results of several numerical studies which support the conclusions of the main article. Furthermore, we provide supplemental information on the experiments described in the main article. The figures of this Supplemental Material are referred to as "Fig. Sx".

## I. NUMERICAL SIMULATIONS

### A. Comparison between experiment and simulation

To estimate the effects of experimental imperfections on the transmission spectra and compare with the experimental results shown in Fig. 4, we performed numerical simulations by a commercial finite-element method solver (COMSOL MULTIPHYSICS). The model used in this simulation is shown in Fig. S1. The unit cell is filled with a lossless dielectric material whose anisotropic dielectric permittivity tensor $\bar{\bar{\epsilon}} = \mathrm{diag}[(n_{\mathrm{Sa}})^2, (n_{\mathrm{Sa}})^2, (n_{\mathrm{Sa}}^{\perp})^2]$, where $n_{\mathrm{Sa}} = 3.1$ and $n_{\mathrm{Sa}}^{\perp} = 3.4$ [1]. Strictly speaking, Babinet's principle does not hold for anisotropic media such as sapphire. However, for the case of normal incidence, the c-cut sapphire substrate can be considered approximately as an isotropic medium. Metal is assumed to be infinitely thin perfect electric conductor (PEC) for simplicity, and resistive sheets are expressed as sheet impedance $Z_{\mathrm{s}}$. It is inevitable in the "sandwich" measurement that the surface of the checkerboard is incompletely covered by the dielectric plate. There should be an air gap between the checkerboard and the cover plate. To incorporate this experimental imperfection into the simulation model, a thin air layer with thickness $t = 2\,\mu\mathrm{m}$ is inserted between the cover plate and the checkerboard screen.

The calculated transmission spectra are shown in Fig. S2(b) for $Z_{\mathrm{s}} = \{10^{-3}, 1, 10^3\} \times Z_0/(2n_{\mathrm{Sa}})$, which correspond to the connected, resistive, and disconnected cases, respectively. The simulations agree with the experimental results qualitatively well in spite of the simplified model.

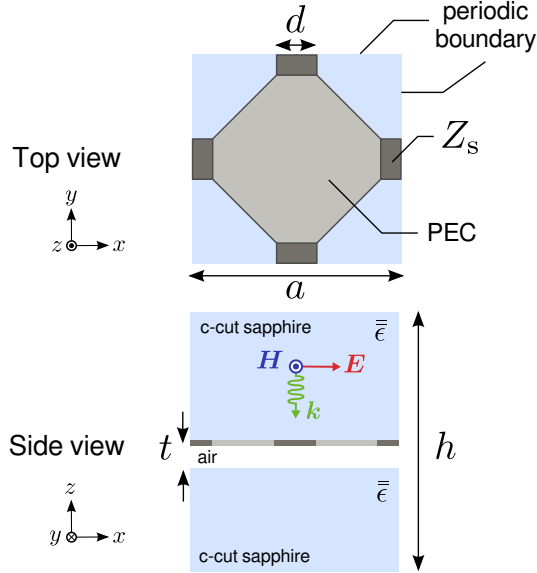

FIG. S1. A schematic picture of the unit cell used in the numerical simulations ($a = 150\,\mu\mathrm{m}$, $h = 450\,\mu\mathrm{m}$, $d = 30\,\mu\mathrm{m}$, and $t = 2\,\mu\mathrm{m}$).

* urade@giga.kuee.kyoto-u.ac.jp
† kitano@kuee.kyoto-u.ac.jp

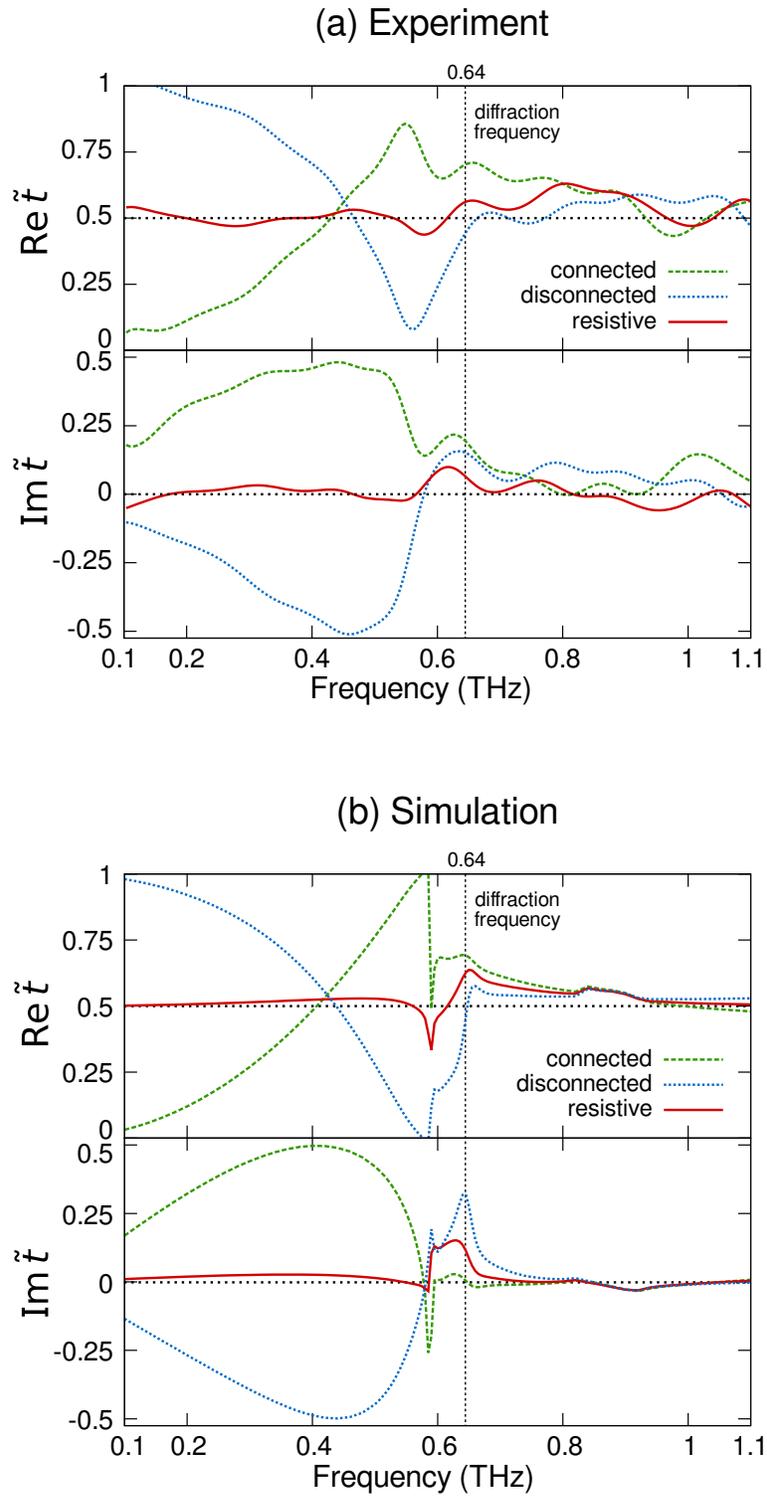

FIG. S2. (a) The measured transmission spectra of the checkerboard screens. (b) The calculated transmission spectra.

## B. Effect of sheet impedance on transmission spectra

Here, we see how the transmission spectrum changes when $Z_\mathrm{s}$ is varied around the self-dual condition $Z_\mathrm{s} = Z_0/(2n_\mathrm{Sa})$. Figure S3 shows the model used in the simulations. For simplicity, the dielectric environment is assumed to be isotropic, and the air gap is not considered. The simulations were performed by ANSYS HFSS. The calculated transmission spectra for $Z_\mathrm{s} = \{4^{-1}, 2^{-1}, 1, 2, 4\} \times Z_0/(2n_\mathrm{Sa})$ are shown in Fig. S4. From the results, it is confirmed that the frequency dependence of the transmission is suppressed as $Z_\mathrm{s}$ approaches the self-dual condition. Although there is a slight deviation from $\tilde{t} = 0.5$ even if $Z_\mathrm{s} = Z_0/(2n_\mathrm{Sa})$, we confirmed that it can be reduced by increasing the mesh density of the model.

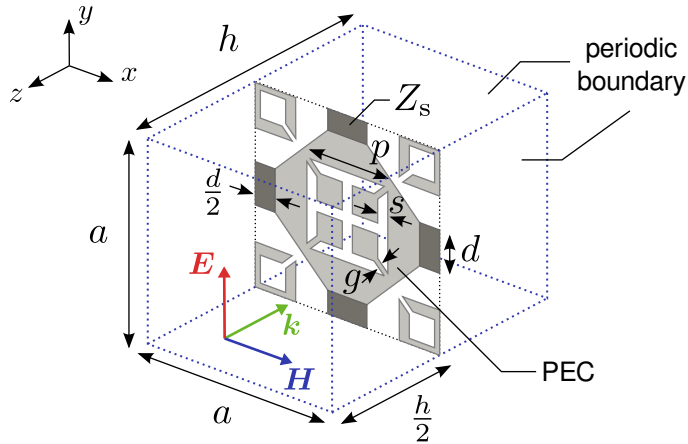

FIG. S3. The model used in the numerical simulations of the checkerboard with the resonant structures ($a = 150\,\mu\mathrm{m}$, $h = 300\,\mu\mathrm{m}$, $d = 30\,\mu\mathrm{m}$, $p = 65\,\mu\mathrm{m}$, $s = 8\,\mu\mathrm{m}$, and $g = 8\,\mu\mathrm{m}$).

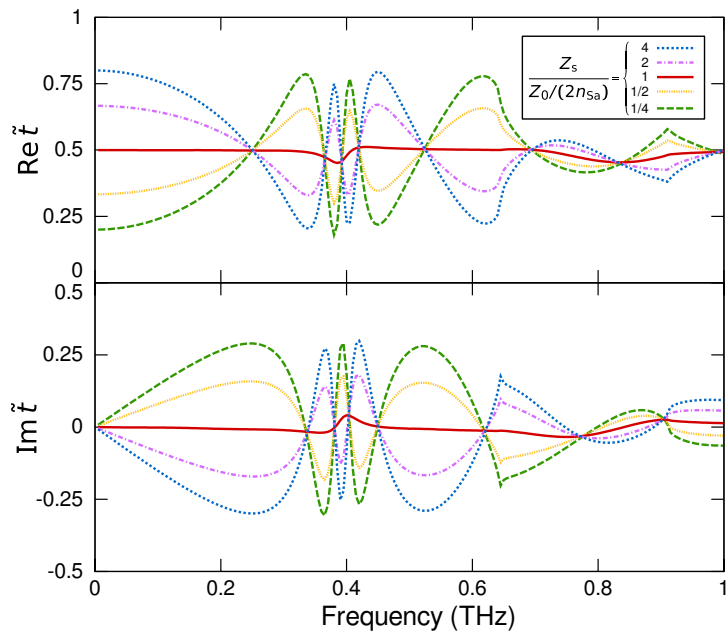

FIG. S4. The calculated transmission spectra of the checkerboard structures with the ELC structures for several values of the sheet impedance.

## C. Resistive structures without self-complementarity

In order to demonstrate the crucial role of self-complementarity, we calculated the transmission spectra of the resistive checkerboard structures without the ELC or complementary ELC (CELC) resonators shown in Figs. S5(a) and S5(b). They are not self-complementary and contain resistive sheets. The model used in this simulation is the same as Fig. S3 except the geometry, and $Z_s$ is set to $Z_0/(2n_{Sa})$. Figure S5(c) shows the calculated transmission spectra. We can confirm that the structures without self-complementarity do not show the frequency-independent response although they contain resistive sheets. This indicates that self-complementarity of the whole structure plays an essential role in realizing the frequency-independent transmission spectrum.

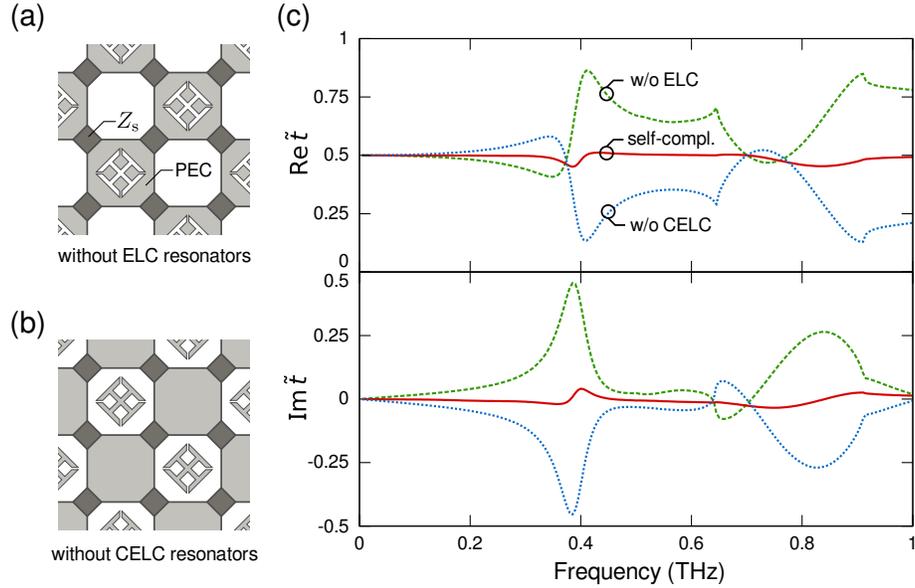

FIG. S5. (a) The resistive checkerboard structure without the ELC resonators. (b) The resistive checkerboard structure without the CELC resonators. (c) The calculated transmission spectra of the resistive checkerboard structures with and without self-complementarity.

## II. SUPPLEMENTAL INFORMATION ON EXPERIMENTS

### A. Experimental setup of THz-TDS

We used the terahertz time-domain spectroscopy (THz-TDS) technique [1] for characterization of the fabricated samples. The experimental setup is as follows. Figure S6 shows a schematic picture of the THz-TDS system. Photoconductive THz antennas (PCAs) on low-temperature-grown GaAs equipped with silicon hyperhemispherical lenses (BATOP) were used as the emitter and detector. These antennas were excited by a femtosecond fiber laser (F-100, IMRA) with a wavelength of 810 nm, through 1.5 m-long optical fibers. The group delay dispersion (GDD) introduced by the optical fibers was compensated by grating pairs [2]. The emitted THz beam was linearly polarized and focused by Teflon lenses on the samples under normal incidence with respect to the sample plane. The beam diameter at the beam waist is estimated as 3.4 mm at 0.5 THz. The transmitted THz electric fields were coherently recorded in the time domain. All the THz components were placed in a box purged with dry air to reduce water vapor absorption of the THz beam.

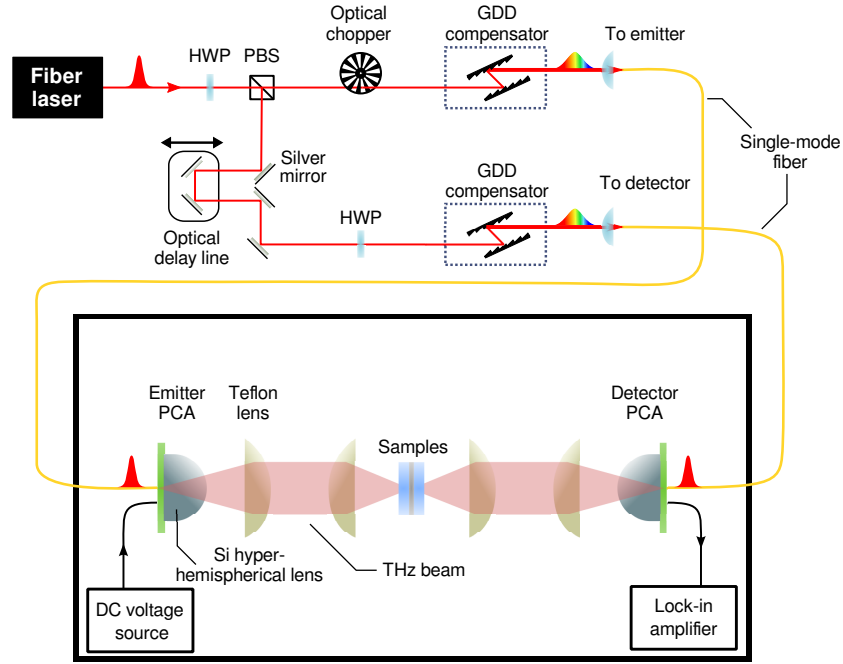

FIG. S6. A schematic picture of the THz-TDS system (HWP: half-wave plate, PBS: polarizing beam splitter).

## B. Transmission spectra of checkerboard screens with resonant structures

Figure S7 shows the transmission spectra which are obtained by Fourier-transforming the temporal waveforms shown in Fig. 5(d). While the connected and disconnected cases exhibit complex spectra due to the interference of multiple resonances, the resistive case shows a relatively flat one around $\tilde{t} = 0.5$ in the observed frequency range. This indicates in the frequency domain that self-complementarity suppresses the frequency dependence of the transmission induced by the ELC resonators.

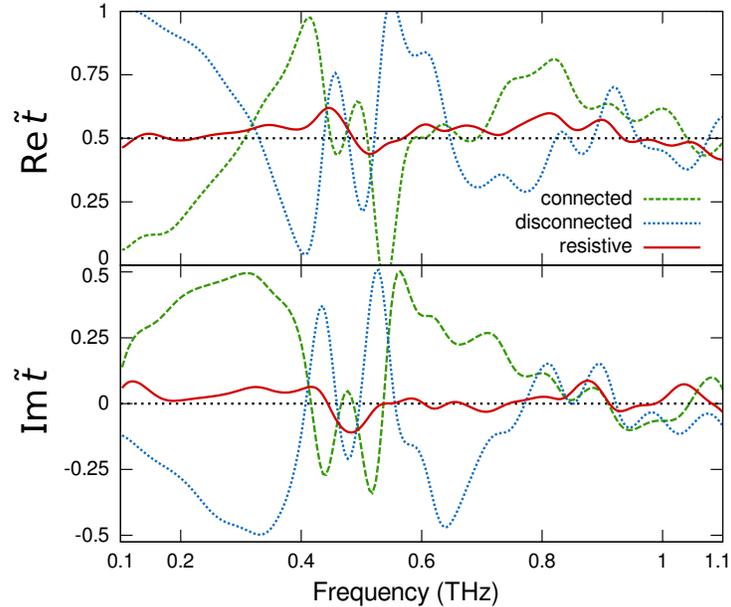

FIG. S7. The measured transmission spectra of the checkerboard screens loaded with ELC resonators.

[1] D. Grischkowsky, S. Keiding, M. van Exter, and Ch. Fattinger, "Far-infrared time-domain spectroscopy with terahertz beams of dielectrics and semiconductors," J. Opt. Soc. Am. B **7**, 2006 (1990).
[2] S. A. Crooker, "Fiber-coupled antennas for ultrafast coherent terahertz spectroscopy in low temperatures and high magnetic fields," Rev. Sci. Instrum. **73**, 3258–3264 (2002).


\end{document}